\documentclass{article}
\usepackage{spconf,amsmath,graphicx,hyperref}
\usepackage[table,xcdraw]{xcolor}
\usepackage{lipsum}
\usepackage{tikz}
\usetikzlibrary{automata,positioning,shapes,backgrounds,fit,calc}
\usetikzlibrary{shapes.geometric, decorations.pathreplacing, arrows.meta}

\usepackage[bottom]{background}
\SetBgContents{To appear in Proc. ICASSP 2026, May 4 - 8, 2026.}
\SetBgVshift{1.0cm}
\SetBgHshift{0.0cm}
\SetBgColor{red}
\SetBgOpacity{0.5}
\SetBgAngle{0.0}
\SetBgScale{1.0}

\newcommand{\rhoi}{\ensuremath{\rho_{I, j}}}
\newcommand{\rhoa}{\ensuremath{\rho_{G, j}}}
\newcommand{\rhoc}{\ensuremath{\rho_{C, j}}}
\newcommand{\versgap}{\ensuremath{v_j}}
\newcommand{\concgap}{\ensuremath{c_j}}
\newcommand{\vspacetop}{\vspace{-1.5mm}}
\newcommand{\vspacebottom}{\vspace{-2.5mm}}

\title{Unseen but not Unknown: Using Dataset Concealment to Robustly Evaluate Speech Quality Estimation Models}
\name{Jaden Pieper and Stephen D. Voran}
\address{Institute for Telecommunication Sciences, Boulder, Colorado, USA \\
\{jpieper, svoran\}@ntia.gov}
\begin{document}

\ninept
\maketitle
\begin{abstract}
We introduce Dataset Concealment (DSC), a rigorous new procedure for evaluating and interpreting objective speech quality estimation models.
DSC quantifies and decomposes the performance gap between research results and real-world application requirements, while offering context and additional insights into model behavior and dataset characteristics.  
We also show the benefits of addressing the corpus effect by using the dataset Aligner from AlignNet when training models with multiple datasets.
We demonstrate DSC and the improvements from the Aligner using nine training datasets and nine unseen datasets with three well-studied models: MOSNet, NISQA, and a Wav2Vec2.0-based model.
DSC provides interpretable views of the generalization capabilities and limitations of models, while allowing all available data to be used at training.
An additional result is that adding the 1000 parameter dataset Aligner to the 
94 million parameter Wav2Vec model during training does significantly improve the resulting model's ability to estimate speech quality for unseen data.

\end{abstract}
\vspacetop
\section{Introduction}
\label{sec:intro}
\vspacebottom
Objective estimation of speech quality is an important and enduring problem. Full-reference estimators (or models) are suitable for out-of-service applications, but only no-reference (NR) models can provide the real-time, in-service results needed for detecting, tracking, and diagnosing the effects of acoustic environments, network loading, and radio conditions in today's complex and dynamic telecommunications environments. NR estimation is a hard problem.  A successful model must  implicitly embody a very general and flexible yet highly detailed representation of what speech should sound like.
Machine learning (ML) has made this hard problem somewhat easier --- ML can leverage a sufficient quantity of speech-quality data (typically speech files and corresponding speech quality values from subjective tests) to train an algorithm so that it can then process previously unseen streams of speech and generate estimates of the corresponding speech quality values.

There is a gap between this very natural, dataset-driven training paradigm and real-world use cases involving arbitrary speech and impairments.
ML-based models typically perform worse on data outside of their training datasets~\cite{Cooper2022}.
In order to improve model performance for real-world applications, we need a clear understanding of how and why these performance drops occur.
We provide a path to developing more useful, robust models via a rigorous evaluation of the relationship between model architecture and training data.
This approach quantifies and decomposes the gap between dataset specific research results and real-world applications.

The primary contribution of this paper is Dataset Concealment (DSC).
DSC is a set of steps for training and evaluating a model that provides a truly interpretable view of its generalization capabilities. 
DSC trains with multiple datasets which increases the range of conditions seen at training and allows us to understand the interactions between datasets and a given model architecture.
Increasing the range of conditions during training increases the ultimate performance on unseen data.
But using multiple datasets for training comes with a cost due to the non-absolute nature of subjective test ratings: the corpus effect, or range-equalizing bias, as demonstrated in~\cite{cooperIS23}.
We address this by using the dataset Aligner module from the AlignNet architecture~\cite{pieper24_interspeech} to learn dataset alignments.
A secondary contribution of this paper is demonstrating that Aligners improve a model's ability to estimate quality at inference for unseen data.

In Section~\ref{sec:bg} we catalog and discuss existing training and evaluation strategies for designing generalizable speech quality estimators.
We also describe the corpus effect and an existing approach to mitigating its impact on model performance.
In Section~\ref{sec:dsc} we formally define DSC.  
Sections~\ref{sec:data} and \ref{sec:results} describe 18 datasets, 3 well-known speech quality models, and the results obtained when DSC is applied using 9 of the datasets. 
Finally, in Section~\ref{sec:inference} we discuss quality estimation results on nine additional unseen datasets for models trained with and without dataset Aligners.

\vspacetop
\section{Background and Motivation}
\label{sec:bg}
\vspacebottom

Developing generalizable speech quality models requires robust and careful evaluation strategies.
One existing strategy is to ``hold out'' randomly selected portions of data at training and to use these for testing. This is a good first step --- any given test file is indeed unseen at training time.  
But the structure of many datasets means that random splitting of files will cause conditions (e.g., street noise at 10 dB SNR), talkers, or sentences to appear in both the training and testing sets.
Cross-validation can be viewed as repeating this operation multiple times in a structured way. While this is a more robust approach, it does not account for the existence of overlap.

Another strategy is to curate splits within datasets so that specific conditions appear only during training and other conditions appear only at testing. This is sensitive to the way the conditions are chosen and begs questions like ``is street noise at 0 and 20 dB SNR different conditions or the same condition?'' An additional drawback is the possibility of other types of overlaps between the training and testing sets, e.g., talkers, speech activity structure, or recording conditions.

A third strategy is to hold out one or more entire datasets as unseen at training and then use model performance on the unseen dataset(s) as a measure of generalization. This should minimize any overlap, but meaningful interpretation of the results can be hindered by two unknowns.  First, how ``easy'' or ``hard'' is an unseen dataset?  A dataset can be easy when a large portion of the label variance is driven by an easily detected condition, such as the low-pass filtering that is used to form anchor conditions in some subjective tests. A dataset can be hard when the majority of the label variance is driven by acoustic nuances such as those that distinguish neural codecs or voice conversion systems that are of similar quality. The second issue for interpreting results on held-out datasets is the question of how different a dataset is from those used in training.  The terms ``in/out of distribution'' and ``in/out of domain'' \cite{scoreq,kibriaIS25,shiIS25} or ``matched/mismatched case'' \cite{de_oliveiraIS25} are used to describe dataset uniqueness in a very coarse and highly subjective way. A more objective and quantitative perspective on dataset uniqueness is needed. 

As will be demonstrated, DSC addresses these issues to provide rich context for understanding how a model performs on a given dataset.
This paper also addresses the corpus effect, which must be considered
when multiple datasets are used during training.
The corpus effect describes the non-absolute nature of MOS ratings that can cause inherent mismatches in subjective labels between different datasets. 
For example, a speech file that received an average rating of 3.0 in one listening experiment could have an average of 3.8 when included in a second experiment because of different contexts, listeners, and experimental designs.
This mismatch between experiments becomes label noise when training a model that must be accurate across multiple datasets. 
Seeking to efficiently use as much data as possible while training,
we explore addressing the corpus effect through the addition of a dataset Aligner~\cite{pieper24_interspeech} during training.
We specifically aim to understand the effect of learning dataset alignments on the inference performance for truly unseen datasets.

\vspacetop
\section{The Dataset Concealment Process}
\label{sec:dsc}
\vspacebottom
DSC is a new combination of training, testing, and analysis steps for $N$ datasets that yields meaningful insights on model performance as well as the datasets in use.
We assume that  each dataset $D_j$ comprises a training set,  a validation set, and a held-out test set.
DSC applied to $N$ datasets entails training a model multiple times to create the following:  
\begin{itemize}
    \item Individual Models: Train the model $N$ times, using each dataset individually, $T_{I, j} = \{D_j\}$
    \item Global Model: Train the model one time using all $N$ datasets together, $T_G = \{D_i\}_{i=1}^N$
    \item Concealed Models: Train the model $N$ times, each time using all datasets except one, $T_{C, j} = \{D_i\}_{i\neq j}$ 
\end{itemize}
The final step of DSC is to apply three different versions of the model to each held-out test set, yielding $3N$ total test results.  
That is,  for dataset $D_j$: (1) Test the Individual Model that was trained using only dataset $D_j$, (2) Test the Global Model that was trained using all datasets,  (3) Test the Concealed Model that was trained using all datasets except $D_j$.
Each testing step produces a figure of merit,
and this could be any figure of merit that quantifies the model's agreement with the subjective labels, e.g., linear correlation coefficient (LCC) or Spearman's rank correlation coefficient (SRCC).
Whenever multiple datasets are used, one must consider the corpus effect which can significantly limit the usefulness of RMSE, especially when a model must produce estimates for previously unseen datasets.
Thus the following definitions are intended for LCC and SRCC and our examples use LCC only, without loss of generality.

We denote the dataset concealment Individual, Global and Concealed LCC values for dataset $D_j$ with $\rhoi, \rhoa,$ and $\rhoc,$ respectively.
\rhoi{} is the correlation observed for the Individual Model that is trained using only dataset $D_j$.
It describes how well the model can learn this dataset in isolation, which is typically the easiest way to learn a dataset.
$\rhoa$ is the correlation observed from the Global Model that is trained with all $N$ datasets $T_G$.
This describes the broad performance of the model, or how well it can combine information from multiple datasets.
The versatility gap of a model on dataset $D_j$ is
\begin{equation}
    \versgap = |\rhoi| - |\rhoa|.
\end{equation}
Currently, when models are trained on multiple datasets, they generally perform worse on any given dataset when compared to individual training.  
So, versatility gap values are generally positive, and the goal is to minimize this gap.
Ideally, a model would be able to harmonize learning from multiple datasets to produce a more  generalized and robust relationship between speech and labels, increasing model performance on any  dataset, and giving negative values of $\versgap$.

The correlation $\rhoc$ is observed from the Concealed Model for dataset $D_j$, where all the datasets except for $D_j$ are used during training.
This correlation describes the model's ability to generalize information from other training datasets for this specific dataset and make estimations on the truly unseen dataset $D_j$.
The concealment gap of a model on dataset $D_j$ is
\begin{equation}
    \concgap = |\rhoa| - |\rhoc|.
\end{equation}
Models that generalize better have smaller concealment gaps.
\begin{figure}[h]
    \centering
    \resizebox{\linewidth}{!}{
        \definecolor{concealedcolor}{HTML}{AA4499}
\definecolor{allcolor}{HTML}{88CCEE}
\definecolor{individualcolor}{HTML}{332288}
\tikzset{
    rhoc/.style={
        rectangle, very thin, fill=concealedcolor, text=white, draw=black!100, font=\footnotesize, minimum width=3mm, minimum height=15 mm,
        },
    rhoa/.style={
        rectangle, very thin, fill=allcolor, text=white, draw=black!100, font=\footnotesize, minimum width=3mm, minimum height=20 mm,
        },
    rhoi/.style={
        rectangle, very thin, fill=individualcolor, text=white, draw=black!100, font=\footnotesize, minimum width=3mm, minimum height=20 mm,
        },
    textlabel/.style={
        font=\small, minimum width=10mm, 
    },
    mathlabel/.style={
         font=\tiny, minimum width=10mm, 
    },
    >=latex,
}

\begin{tikzpicture}[ultra thick, node distance=6mm]
    \node (rhoc1) [rhoc, alias=last] {\rotatebox{90}{Concealed}};
    \node (pc1) [mathlabel, above=-2mm of last.north, xshift=-5mm] {\rhoc};
    \node (rhoa1) [rhoa, right=of last.south, anchor=south, minimum height=23 mm, alias=last] {\rotatebox{90}{Global}};
    \node (pa1) [mathlabel, above=-2mm of last.north, xshift=-6.5mm] {\rhoa};
    \node (ci) [mathlabel, left=1.5mm of rhoa1.north, yshift=-4mm] {\concgap};
    \path[-,draw,decorate,decoration={brace, mirror, raise=2.1pt, amplitude=6pt}] ([xshift=-1.8mm]rhoa1.north) -- ([xshift=4.2mm]rhoc1.north);
    \node (rhoi1) [rhoi, right=of last.south, anchor=south, minimum height=30 mm, alias=last] {\rotatebox{90}{Individual}};
    \node (pi1) [mathlabel, above=-2mm of last.north, xshift=-6.5mm] {\rhoi};
    \node (vi) [mathlabel, left=1.5mm of rhoi1.north, yshift=-3.8mm] {\versgap};
    \path[-,draw,decorate,decoration={brace, mirror, raise=2.1pt, amplitude=6pt}] ([xshift=-1.8mm]rhoi1.north) -- ([xshift=4.2mm]rhoa1.north);
    \node (model1) [textlabel, below=0mm of rhoa1.south,] {Model 1};
    \node (rhoc2) [rhoc, right=20mm of last.south, anchor=south, alias=last] {\rotatebox{90}{Concealed}};
    \node (rhoa2) [rhoa, right=of last.south, anchor=south, minimum height=28 mm, alias=last] {\rotatebox{90}{Global}};
    \node (rhoi2) [rhoi, right=of last.south, anchor=south, minimum height=30 mm, alias=last] {\rotatebox{90}{Individual}};
    \node (model2) [textlabel, below=0.5mm of rhoa2.south,] {Model 2};
    \node (rhoc3) [rhoc, right=20mm of last.south, anchor=south, minimum height=27 mm, alias=last] {\rotatebox{90}{Concealed}};
    \node (rhoa3) [rhoa, right=of last.south, anchor=south, minimum height=28 mm, alias=last] {\rotatebox{90}{Global}};
    \node (rhoi3) [rhoi, right=of last.south, anchor=south, minimum height=30 mm] {\rotatebox{90}{Individual}};
    \node (model3) [textlabel, below=0.5mm of rhoa3.south,] {Model 3};
\end{tikzpicture}
    }
    \vspace{-5mm}\caption{
    Example dataset concealment results and interpretation. 
    }
    \label{fig:dsc-example}
\end{figure}
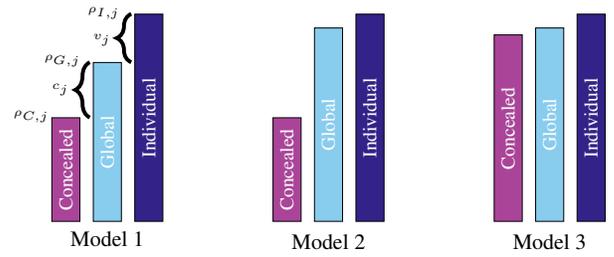
\vspace{-2mm}

 Figure~\ref{fig:dsc-example} provides a pictorial representation of the three correlations and the two gaps for three models and one dataset.
The figure shows a progression of model improvement (Model 1 to Model 2 to Model 3).  The three models show similar performance, $\rhoi$, when trained on the dataset individually. 
But Model 1 is not very versatile at learning from multiple datasets at once, as seen through its significant versatility gap; when trained with multiple datasets, it loses the ability to perform well on the considered dataset.
Model 2 reduces the versatility gap,
but does not generalize well.  This is clear because it has a significant concealment gap --- when the considered dataset is concealed during training, Model 2 does poorly on that dataset.
Model 3 has minimal versatility and concealment gaps so we conclude that Model 3 is both versatile and generalizable.

\vspacetop
\section{Datasets and Models}
\label{sec:data}
\vspacebottom
Table \ref{tab:data} summarizes key attributes of the nine datasets used to demonstrate DSC and nine other datasets that are truly unseen and used for additional context on generalization ability of models. The NOIZEUS and PSTN datasets contain narrowband speech (nominal upper limit at 3.4 kHz) and the remaining datasets contain wideband speech (7 kHz limit) or fullband speech (20 kHz limit).

\begin{table}[t]
    \centering
    \resizebox{\linewidth}{!}{
        \begin{tabular}{|l|l|r|l|r|l|} \hline
\textbf{Dataset}                 & \textbf{Year} & \textbf{Files} & \textbf{Conditions} & \begin{tabular}{c} \textbf{Votes} \\ \textbf{per} \\ \textbf{File} \end{tabular} & \textbf{Use} \\ \hline
\begin{tabular}{l}FFTNet \end{tabular} \cite{JinICASSP2018}                 & 2018 & 1200     & ~~Neural codecs & 8.9          &      DSC      \\ \hline
\begin{tabular}{l} NOIZEUS  \cite{Loizou2006} \end{tabular}                & 2007 & 1664     & ~~Noise \& NR  & 8.0         &   DSC      \\ \hline
\begin{tabular}{l} VMC2022   \cite{voicemos2022} \end{tabular}               & 2022     & 7106       & ~~Synthesized \& natural speech           &        8.0        &    DSC        \\ \hline
\begin{tabular}{l} Tencent  \cite{ConfSpeech2022} \end{tabular}               &  2022    & 11,563   &   \begin{tabular}{l} Noise, NR, reverb, codecs, \\ packet loss, PLC \end{tabular}        &      20.0          &    DSC        \\ \hline
\begin{tabular}{l} NISQA \\ SIM \cite{Mittag2021IS} \end{tabular}             &  2021    & 12,500     &   \begin{tabular}{l} Codecs, packet loss, noise, \\ filtering, clipping \end{tabular}        &      5.2          &   DSC         \\ \hline
\begin{tabular}{l} TMHINT-QI  \cite{TMHINT} \end{tabular}        &  2022    & 14,915     &      ~~Noise \& NR      &   1.6              &     DSC       \\ \hline
\begin{tabular}{l} VCC2018\textdagger\cite{VCC2018} \end{tabular}                 &  2018    & 19,670     & ~~Voice conversion algorithms           &      4.0          &   DSC         \\ \hline
\begin{tabular}{l} IU    \cite{Dong2020}  \end{tabular}                  &  2020    & 36,000     &  ~~Noise \& reverb          &   5.0             &    DSC        \\ \hline
\begin{tabular}{l} PSTN   \cite{Mittag2020Interspeech} \end{tabular}                  &  2020    & 58,709     &     ~~PSTN/VoIP calls, noise       &     4.6           &    DSC        \\ \hline \hline
\begin{tabular}{l} NISQA  \\ Livetalk  \cite{Mittag2021IS} \end{tabular}        &  2021    & 232      &   \begin{tabular}{l} Live calls (wired  \& wireless,\\ PSTN \& VoIP), talking levels, \\natural and recorded noise \end{tabular}   &      24.0          &    Unseen      \\ \hline
\begin{tabular}{l} NISQA \\ FOR  \cite{Mittag2021IS} \end{tabular}             &   2021   & 240      &   \begin{tabular}{l} Codecs, noise, packet loss, \\clipping, live OTT calls \end{tabular}        &         29.3       &   Unseen       \\ \hline
\begin{tabular}{l} NISQA \\ NSC   \cite{Mittag2021IS} \end{tabular}           &  2021    & 240      &   \begin{tabular}{l} Codecs, noise, packet loss, \\ clipping \& live calls \end{tabular}       &    27.2            &   Unseen       \\ \hline
\begin{tabular}{l} NISQA \\ P501   \cite{Mittag2021IS} \end{tabular}           &  2021    & 240      &  \begin{tabular}{l} Codecs, noise, packet loss, \\ clipping \& live calls \end{tabular}          &     28.3           &   Unseen       \\ \hline
\begin{tabular}{l} Blizzard2021 \\ SS1 Nat. \cite{Blizzard2021} \end{tabular}  &  2021    & 242       &  ~~Synthesized \& natural speech         &    16.0            &   Unseen       \\ \hline
\begin{tabular}{l} Blizzard2021 \\ SH1 Nat. \cite{Blizzard2021} \end{tabular}  &  2021    & 338      & ~~Synthesized \& natural speech           &      24.1          &   Unseen       \\ \hline
\begin{tabular}{l} Blizzard2021 \\ SS1 Acc. \cite{Blizzard2021}  \end{tabular} &   2021   & 363       &   ~~Synthesized \& natural speech         &     16.1           &   Unseen       \\ \hline
\begin{tabular}{l} Blizzard2008 \\ News Nat. \cite{Blizzard2008} \end{tabular}      &    2008  & 802    &  ~~Synthesized \& natural speech          &    10.5            &  Unseen        \\ \hline
\begin{tabular}{l}Blizzard2008 \\ Novel  Nat. \cite{Blizzard2008} \end{tabular}     &   2008   & 802    &   ~~Synthesized \& natural speech         &        10.1        &    Unseen      \\ \hline
\end{tabular}
    }
    \caption{
    Summary of datasets used. 
    ``Overall speech quality" was rated except where ``Nat." indicates that ``speech naturalness" was rated and ``Acc." means that ``speech acceptability" was rated.
    NR: noise reduction, PLC: packet loss concealment, OTT: over-the-top. 
    \textdagger VCC2018 has been punctured to remove overlap with VMC2022.
    }
    \label{tab:data}
    \vspace{-5mm}
\end{table}

We demonstrate DSC for three different well-established and well-studied speech quality models. 
This allows us to compare what DSC tells us about these models with what is already understood about these models. 
The intended use case for DSC is the evaluation of proposed \textit{new} models, thoroughly gathering rich, well-contextualized insights that enable us to see strengths and weaknesses relative to other models.  
It is our intent to do this going forward, and we invite other researchers to leverage DSC as well.

The first model we consider is MOSNet~\cite{Lo2019}.
This model represents an early approach to NR ML estimation that relies on a CNN and BLSTM based architecture.
It is a medium-sized model with 1.4 million parameters and is the basis for subsequent work~\cite{Leng2021,Huang2022}.
MOSNet is known for poor generalization, so it provides a useful example model for DSC.
We trained MOSNet as described in~\cite{pieper24_interspeech}.

The second model we consider is a single-headed NISQA~\cite{mittag2021nisqa}, as most of the datasets used here have labels for speech quality only.
NISQA is an early attention-based approach to speech quality estimation that is extremely light-weight and successful and is known to generalize well~\cite{ConfSpeech2022}, despite  having only 218,000 parameters.

Our third example is a Wav2Vec2.0-based~\cite{Baevski-2020-wav2vec2.0} speech quality model, often called SSL-MOS~\cite{Cooper2022}, but here referred to as Wav2Vec, for simplicity.
This model is an early effort to leverage massive self-supervised speech models for speech quality estimation.  This is done by attaching a single fully-connected layer to the feature output from the SSL model.
The Wav2Vec model has 94 million parameters.
Large SSL-based models are known to be extremely robust and generalizable at inference for unseen data.

We trained Global and Concealed Models in two different ways: conventionally and with the addition of an Aligner as in the AlignNet architecture~\cite{pieper24_interspeech}.
Conventional training treats all datasets as one, despite the fact that their labels may not be compatible.
The AlignNet architecture appends a small and effective dataset Aligner to the output of the model to mitigate the corpus effect between datasets. 
The corpus effect can cause incompatibility between labels from one subjective test and those from a different subjective test.
The Aligner uses a dataset indicator to map an intermediate, reference dataset score to the appropriate target dataset scale, effectively learning the alignment functions between datasets.
This reduces the dissonance involved in learning from multiple datasets, and allows the audio model to focus solely on accurately learning the audio space.

ML is a powerful tool and some models are able to do some learning through the label noise introduced by the corpus effect, without the use of an Aligner~\cite{huang2024}.
However, removing label noise during the training process
allows for the learning of more robust relationships between speech and quality and thus
improves model performance at inference on unseen data.
We expect the magnitude of this effect from the Aligner to be model-dependent.

\vspacetop
\section{Example Dataset Concealment Results}
\label{sec:results}
\vspacebottom
\begin{figure*}[h]
    \centering
    \includegraphics[width=\linewidth]{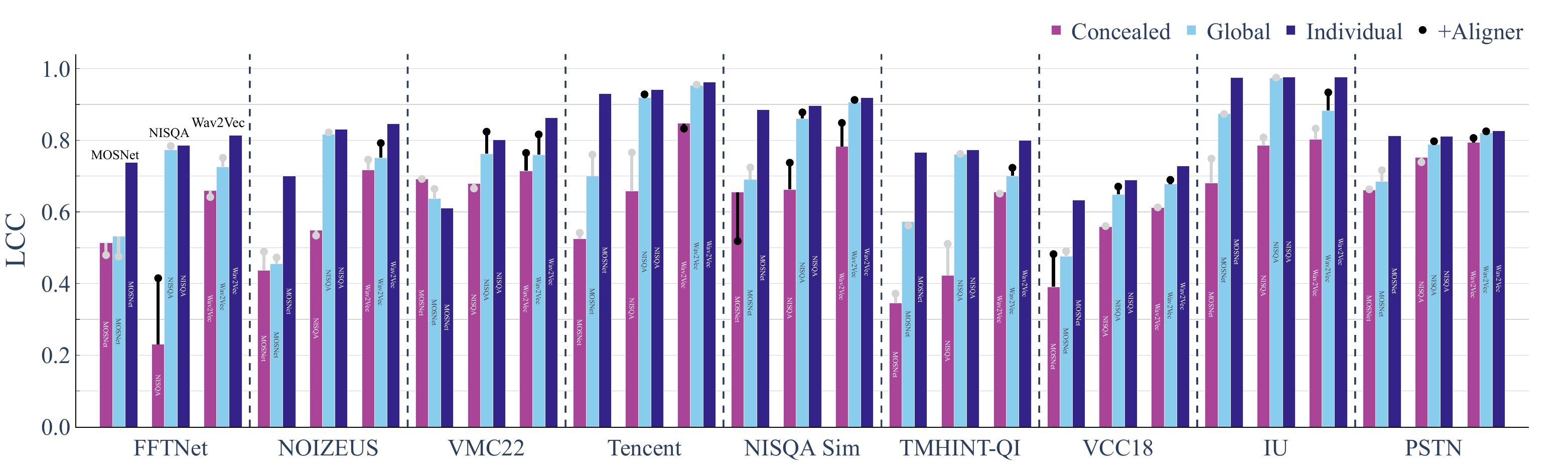}
    \caption{DSC results for MOSNet, NISQA, and Wav2Vec models across nine datasets. Bars show performance when training conventionally and lines extending from bars show the effect of training with an Aligner. Black lines indicate statistically significant changes, grey otherwise. }
    \label{fig:dsc-results}
    \vspace{-5mm}
\end{figure*}

We ran a variety of experiments to demonstrate the insights that DSC provides and the improvements that training with an Aligner can offer.
We trained Global and Individual models 10 times each with randomized training/validation/testing splits, respecting curated splits where applicable.
Due to the number of datasets used here, we ran two replications when training Concealed Models.
We report ``average’’ correlations (we average the Fisher z-transformation\cite{fisher1915} of the per-replication correlations and then apply the inverse of the Fisher z-transformation).
Statistical significance is determined by computing the standard error of the Fisher z-transformed average correlation and computing a 95\% confidence interval of the difference of the transformed averages.
Training details specific to dataset alignment include: 
NISQA Sim is  the reference dataset; 
Tencent is the reference dataset when NISQA Sim is the concealed dataset;
when training MOSNet and NISQA the Aligner is frozen until the validation correlation reaches a threshold of 0.6, so that the model can reach a meaningful representation before the learning of dataset alignments; 
when training Wav2Vec, due to the existing SSL-based speech representation and comparatively long training times, the Aligner is never frozen.
Full details are provided at \url{https://github.com/NTIA/Dataset-Concealment}.

\begin{figure}
    \centering
    \includegraphics[width=\linewidth]{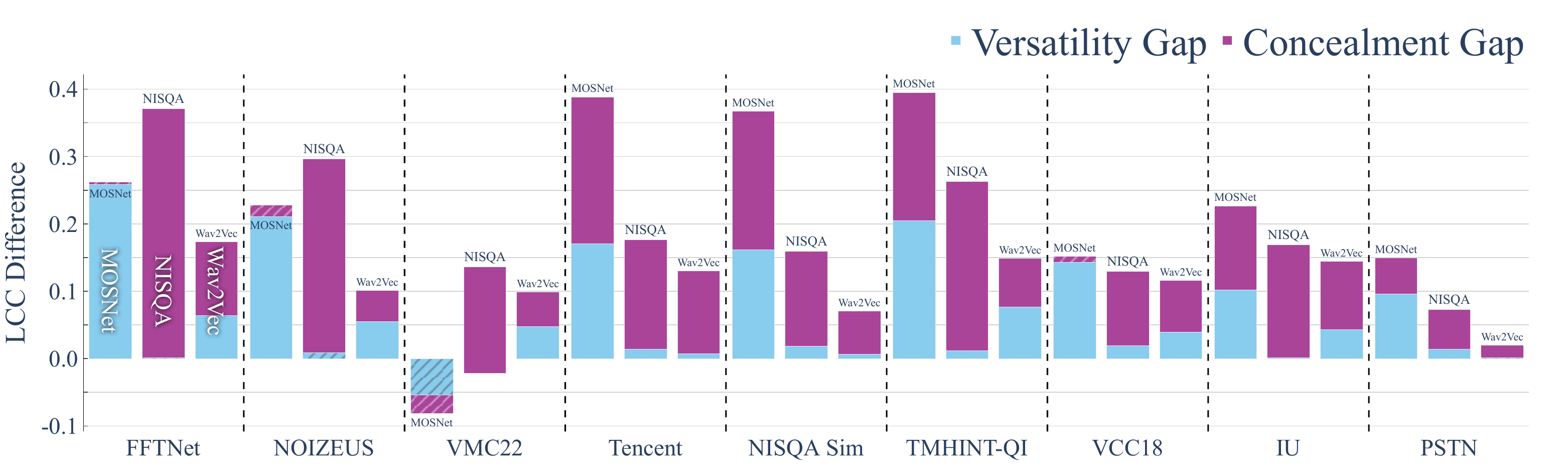}
    \caption{DSC versatility and concealment gaps for models trained with an Aligner. Solid bars represent statistically significant gaps, otherwise bars are hatched.}
    \label{fig:dsc-gaps}
    \vspace{-5mm}
\end{figure}
The DSC LCC results and the impact of adding an Aligner during training can be seen in Fig.~\ref{fig:dsc-results}. Figure~\ref{fig:dsc-gaps} shows the
DSC versatility gaps $v_i$ and concealment gaps $c_i$, but only for models trained with an Aligner, due to space constraints.
Insights about models, datasets, and interactions between the two can be easily seen in these two figures.
For Individual Models, MOSNet performs worse than NISQA, which performs worse than Wav2Vec.
This is well known and expected, and often where analysis of this sort starts and ends; but DSC provides much more information.

The Global Models show a different trend.
Figure~\ref{fig:dsc-gaps} makes it clear that, both conventionally and with an Aligner, NISQA has very low versatility gaps --- training NISQA with multiple datasets has a very small impact on its ability to predict any individual dataset.
Wav2Vec with an Aligner, on the other hand, has versatility gaps that are slightly larger than those of NISQA, indicating that Wav2Vec has slightly lower versatility.
In fact, Fig.~\ref{fig:dsc-results} shows that the NISQA Global Model has higher correlations than the Wav2Vec Global Model on the FFTNet, NOIZEUS, VMC22, TMHINT-QI, and IU datasets.
MOSNet shows the largest versatility gaps for every dataset outside of VMC22, and  the MOSNet $\rho_G$ values in Fig.~\ref{fig:dsc-results} show that MOSNet struggles to learn from multiple datasets at all.
This is different behavior from what was observed in~\cite{pieper24_interspeech}, which we posit is due to using better defined, more challenging splits for the datasets and the group of DSC training datasets being more challenging and diverse than the two groups used in that work.

Adding an Aligner does not produce any statistically significant differences in MOSNet  Global Models.
But adding an Aligner to NISQA Global Models during training produces statistically significant LCC increases for five out of nine datasets.
And adding an Aligner to Wav2Vec Global Models gives statistically significant LCC increases for seven out of nine datasets.
Wav2Vec, in many ways,
can be considered a state-of-the-art speech quality model, and adding an Aligner improved its ability to predict speech quality in held-out test sets.
The fact that adding a 1000-parameter Aligner to a 94-million parameter model produces a statistically significant improvement in the majority of cases demonstrates that label noise due to the corpus effect is  impacting the ability of a model to learn and this can be efficiently mitigated.

Concealed training results describe how well each model can predict speech quality in a dataset that has not been seen during training.
Figures~\ref{fig:dsc-results} and \ref{fig:dsc-gaps} show that the Wav2Vec concealment gaps are generally the smallest.
The results for PSTN with Wav2Vec are particularly impressive, with Concealed training achieving a correlation of $\rhoc = 0.81$ compared to the Individual training result of $\rhoi = 0.83$.
Figure~\ref{fig:dsc-gaps} makes it clear that this difference is almost entirely due to the Concealment gap --- the versatility gap is not statistically significant.
NISQA models have larger Concealment gaps than Wav2Vec models for almost every dataset.
So while NISQA proves itself to be more versatile than Wav2Vec in terms of training with multiple datasets, Wav2Vec shows a better ability to generalize to unseen data than NISQA. This is an example of the type of nuanced model comparison that only DSC can provide.
Finally, Concealed MOSNet models have a very poor showing, reinforcing MOSNet's limitations as a model that tends to overfit to the data it is trained on.
In general, adding an Aligner has lesser impact in Concealed training than it does in Global training.

DSC also yields insights about datasets.
For example, the FFTNet dataset gives the largest and second largest concealment gap for NISQA and Wav2Vec respectively, consistent with the unique conditions in that dataset.
The Tencent dataset gives the largest MOSNet concealment gap, second largest for Wav2Vec, and fifth largest for NISQA.
While the Tencent dataset shares telecommunication conditions with the NISQA Sim and TMHINT-QI datasets, it apparently is novel in other relevant aspects (e.g., recording conditions, speech material, or talkers).

\begin{figure}[h]
    \centering
    \includegraphics[width=\linewidth]{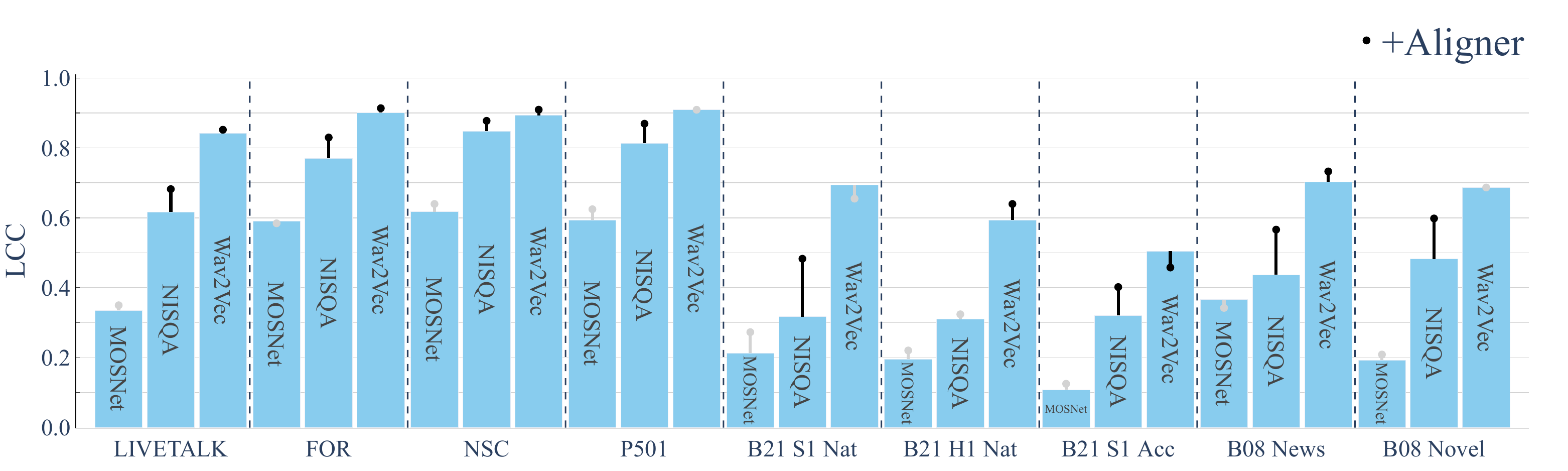}
    \caption{Inference performance of MOSNet, NISQA, and Wav2Vec across nine unseen datasets. Bars show the performance when training conventionally.
    Lines extending from the bars show the effect of training with an Aligner. 
    Black lines indicate statistically significant changes, grey otherwise.
    }
    \label{fig:inf-results}
    \vspace{-5mm}
\end{figure}

\vspace{-7mm}
\section{Inference Results}\label{sec:inference}
\vspacebottom

Figure~\ref{fig:inf-results} shows inference LCC values for globally-trained MOSNet, NISQA, and Wav2Vec, trained with and without an Aligner.  These inference results are for nine unseen test datasets (datasets not used in DSC).
The trends in Fig.~\ref{fig:inf-results} agree with what we have already learned from the DSC concealment gaps: for truly unseen datasets, Wav2Vec performs the best, NISQA is somewhat worse, and MOSNet estimates are unlikely to be useful.
Training with an Aligner proves to be beneficial for the models at inference.
The addition of an Aligner gives statistically significant improvements for eight of the nine datasets for NISQA and five of the nine for Wav2Vec.
Wav2Vec has proved itself to be one of the more powerful current models for producing estimates for unseen data, so the fact that a 1000-parameter Aligner from the AlignNet architecture can improve its inference ability demonstrates the value of acknowledging and attempting to mitigate the impact of the corpus effect during training.

\vspacetop
\section{Conclusion}
\vspacebottom
We have defined DSC, applied it to three speech quality models and nine datasets, and described the insights that it provides.
DSC produces a versatility gap and a concealment gap for each model and dataset. 
These illuminate the limitations of a model architecture and add context that can help inform future architectural designs.
We evaluated the models with nine additional truly unseen datasets, and those results affirm the insights derived from using DSC.
In addition, both sets of experiments demonstrate that adding a lightweight Aligner when training with multiple datasets yields statistically significant improvements most of the time, both for held-out test sets and for truly unseen data.


\bibliographystyle{IEEEbib}
\bibliography{sources}

\end{document}